\documentclass[sigconf]{acmart}

\usepackage{enumitem}
\usepackage{fontawesome}
\usepackage{soul}

\AtBeginDocument{%
  }

\setcopyright{none}
\copyrightyear{2025}
\acmYear{2025}
\acmDOI{}

\newcommand{\codescribe}{CodeScribe}
\newcommand{\mcfm}{MCFM}
\acmConference[PASC]{The Platform for Advanced Scientific Computing}{June 16--18, 2025}{Brugg, Switzerland}
\acmISBN{}

\begin{document}

\title{Leveraging Large Language Models for Code Translation and Software Development in Scientific Computing}

\author{Akash Dhruv}
\affiliation{%
 \institution{Argonne National Laboratory}
 \city{Lemont}
 \state{Illinois}
 \country{USA}}

\author{Anshu Dubey}
\affiliation{%
 \institution{Argonne National Laboratory}
 \city{Lemont}
 \state{Illinois}
 \country{USA}}

\renewcommand{\shortauthors}{Dhruv \& Dubey}
\newcommand\codetext[1]{{\small\texttt{#1}}}

\begin{abstract}
The emergence of foundational models and generative artificial intelligence (GenAI) is poised to transform productivity in scientific computing, especially in code development, refactoring, and translating from one programming language to another. However, because the output of GenAI cannot be guaranteed to be correct, manual intervention remains necessary. Some of this intervention can be automated through task-specific tools, alongside additional methodologies for correctness verification and effective prompt development. We explored the application of GenAI in assisting with code translation, language interoperability, and codebase inspection within a legacy Fortran codebase used to simulate particle interactions at the Large Hadron Collider (LHC). In the process, we developed a tool, \codescribe, which combines prompt engineering with user supervision to establish an efficient process for code conversion. In this paper, we demonstrate how \codescribe~assists in converting Fortran code to C++, generating Fortran-C APIs for integrating legacy systems with modern C++ libraries, and providing developer support for code organization and algorithm implementation. We also address the challenges of AI-driven code translation and highlight its benefits for enhancing productivity in scientific computing workflows.
\end{abstract}



\received{DD-MM-YYY}
\received[revised]{DD-MM-YYY}
\received[accepted]{DD-MM-YYY}

\maketitle

\section{Introduction}

The widespread presence of GPUs in high-performance computing (HPC) platforms and the use of C++ abstractions to simplify GPU programming have spurred significant interest in converting existing Fortran codes to C++. In the Exascale Computing Project (ECP), several scientific applications underwent this transition to adopt Kokkos \cite{kokkos} as their GPU programming model. Most of these conversions were performed manually, taking years of development and translation effort. The advent of generative AI (GenAI) offers substantial potential to reduce the burden of manual code translation.

Translating legacy Fortran codebases to C++ has long been a challenging task for software developers. This difficulty arises not only from the nuances of the Fortran language but also from diverse coding practices accumulated over decades of software evolution. Many scientific applications contain code optimized for specific hardware, which complicates direct translation to C++. Additionally, the need for performance optimization across modern HPC platforms introduces another layer of complexity. Bulk translation of entire codebases is seldom a viable option, as non-trivially complex codebases often encounter broken functionality and degraded performance, leading to lengthy debugging processes that can consume significant developer time.

In practice, manual incremental translation has been the predominant approach in the community. Typically, developers convert code sections gradually, build Fortran-C interfaces, and run tests to ensure correctness. This controlled approach allows developers to verify the functionality of each section before moving on to the next. However, this process is labor-intensive, requiring extensive knowledge of both Fortran and C++ and an understanding of the interoperability challenges between the two languages. We investigated the use of GenAI to aid in translating \mcfm, a Monte Carlo code that simulates particle interactions observed at the Large Hadron Collider (LHC), from Fortran to C++. To this end, we developed a tool, \codescribe, along with a methodology for efficient translation. This paper describes \codescribe~and its application to \mcfm~and also highlights other legacy Fortran codes under consideration.

The paper is organized as follows: Section \ref{sec:background} reviews the literature on code translation and the use of GenAI for developer productivity. Section \ref{sec:codescribe} provides an overview of \codescribe, followed by the translation methodology in Section \ref{sec:methodology}. Section \ref{sec:results} presents results from large language model sensitivity studies, and Section \ref{sec:ongoing-work} discusses ongoing and future work.

\section{Background}
\label{sec:background}
The field of scientific computing has become increasingly reliant on the rapid development and deployment of advanced software tools. As simulations and models drive progress in everything from climate research to high-energy physics, the effective use of large language models (LLMs) in software development is becoming an essential component of the scientific workflow. The application of GenAI in software development, particularly in code translation and refactoring, has gained significant attention in recent years. Numerous studies have explored the potential of language models to automate the translation of legacy programming languages into more modern ones, enhancing interoperability and maintainability. For instance, research by Alon et al. \cite{alon2019code} demonstrated that deep learning models could be employed to generate code snippets based on natural language descriptions, thereby bridging the gap between non-technical stakeholders and software development processes. Additionally, recent advancements in retrieval-augmented generation (RAG) have further refined this approach, allowing models to utilize external knowledge bases to enhance their code generation capabilities (Lewis et al. \cite{lewis2020retrieval}). The RAG framework has been particularly effective in improving the contextual understanding of complex queries, thereby producing more accurate and relevant responses from the model. As generative AI continues to evolve, its integration into scientific computing, especially for translating legacy Fortran code to C++, offers a promising avenue for enhancing developer productivity and driving innovation in high-performance computing environments. While human intervention may still be required to verify correctness of the generated code, the use of AI allows developers to concentrate on higher-level architectural decisions and performance tuning, rather than being slowed down by time-consuming manual tasks.

Recent advancements in specialized large language models (LLMs), such as CodeLlama \cite{Codellama2024} and OpenAI’s Codex \cite{Codex2021}, which powers GitHub Copilot \cite{copilot}, have opened new avenues for automating various software development processes. These models excel at parsing, transforming, and generating code across different programming languages, making them particularly valuable in code generation and code-assist tools. The use of these models via their application programming interfaces (APIs) has enhanced application development by facilitating integration and improving productivity, as evidenced by the success of GitHub Copilot.

Frameworks like LASSI \cite{Dearing2024} further enhance this automation by streamlining prompt generation, compilation, and execution processes. Originally designed to generate synthetic coding data for training GenAI models, LASSI focuses on producing working code, but the assurance of code correctness remains a significant challenge and an unresolved implementation issue. This highlights the need for ongoing developer oversight, even as automation tools advance. 

 We have developed \codescribe~\cite{akash_dhruv_2024_13879406}, an AI-assisted tool designed to facilitate the translation of Fortran codebases to C++ as a step towards integrating GenAI technology in scientific  code development workflows. We explore its application in code translation for scientific computing, and compare the performance and correctness of different LLMs in this context. Additionally, we highlight how \codescribe~has been applied to various scientific computing software projects, for code translation and code inspection to enhance developer productivity. Finally, we outline future directions for enhancing the tool and integrating it into scientific workflows.

\section{Overview of \codescribe}
\label{sec:codescribe}
{ In scientific computing, the urgency to translate legacy Fortran codebases to modern languages like C++ is driven by the desire to leverage advanced libraries and hardware optimizations for HPC platforms. While Fortran remains highly performant, its 
support for platform heterogeneity remains subpar compared to that of C++. Translating legacy codes to C++ makes a much richer ecosystem for performance portability accessible to these codes. However, this translation process is fraught with challenges \cite{CARY199720}}.

A primary difficulty in Fortran-to-C++ translation arises from the fundamental differences between the languages. Fortran is renowned for its array operations and implicit type handling, while C++ demands explicit declarations and offers greater control over memory management. Additionally, large-scale scientific codes written in Fortran often evolve over decades, resulting in inconsistent coding styles, varying degrees of optimization, and numerous implicit assumptions about hardware.

\begin{figure}[h]
    \begin{center}
        \includegraphics[width=\linewidth]{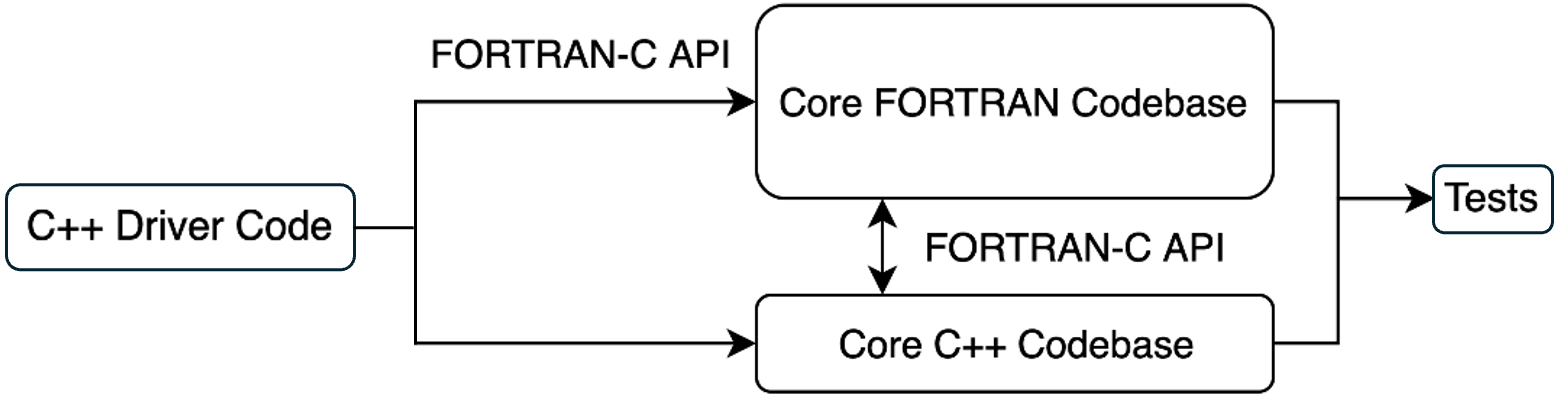}
        \caption{Workflow diagram illustrating the interaction between C++ driver code and the core codebase, with the Fortran-C API facilitating interoperability.}
        \Description{Workflow diagram illustrating the interaction between C++ driver code and the core codebase, with the Fortran-C API facilitating interoperability}
         \label{fig:code-translation-workflow}
    \end{center}
\end{figure}

An alternative method to full-scale rewriting is incremental translation, where code is converted in smaller, manageable sections, creating interfaces (usually Fortran-C layers) between legacy Fortran code and newly written C++ sections. This layered approach preserves functionality as each code segment is tested, facilitating easier debugging by isolating issues in well-defined portions. Figure \ref{fig:code-translation-workflow} provides an example of such a workflow where the driver C++ code interfaces with the core codebase written in both Fortran and C++. Interoperability with the Fortran part of the codebase is facilitated by the Fortran-C API. To ensure a smooth transition, the translation process typically begins with converting data structures, which are foundational to the code's organization, before moving on to methods and functionality. As the translation progresses, more of the Fortran codebase is converted to C++ and verified for correctness by running tests.

Despite its advantages, incremental translation remains labor-intensive. Developers must manually manage Fortran-C layers, handle complex data structures, and ensure that performance optimizations are retained. This is where GenAI can make a significant impact. However, effectively leveraging AI within an application requires a careful understanding of the task at hand, domain insight, and effective prompt development. Through our exploration of this field, we distilled our knowledge into \codescribe, a tool that offers both code translation and software development assistance.

\begin{figure*}[h]
    \begin{center}
        \includegraphics[width=0.62\linewidth]{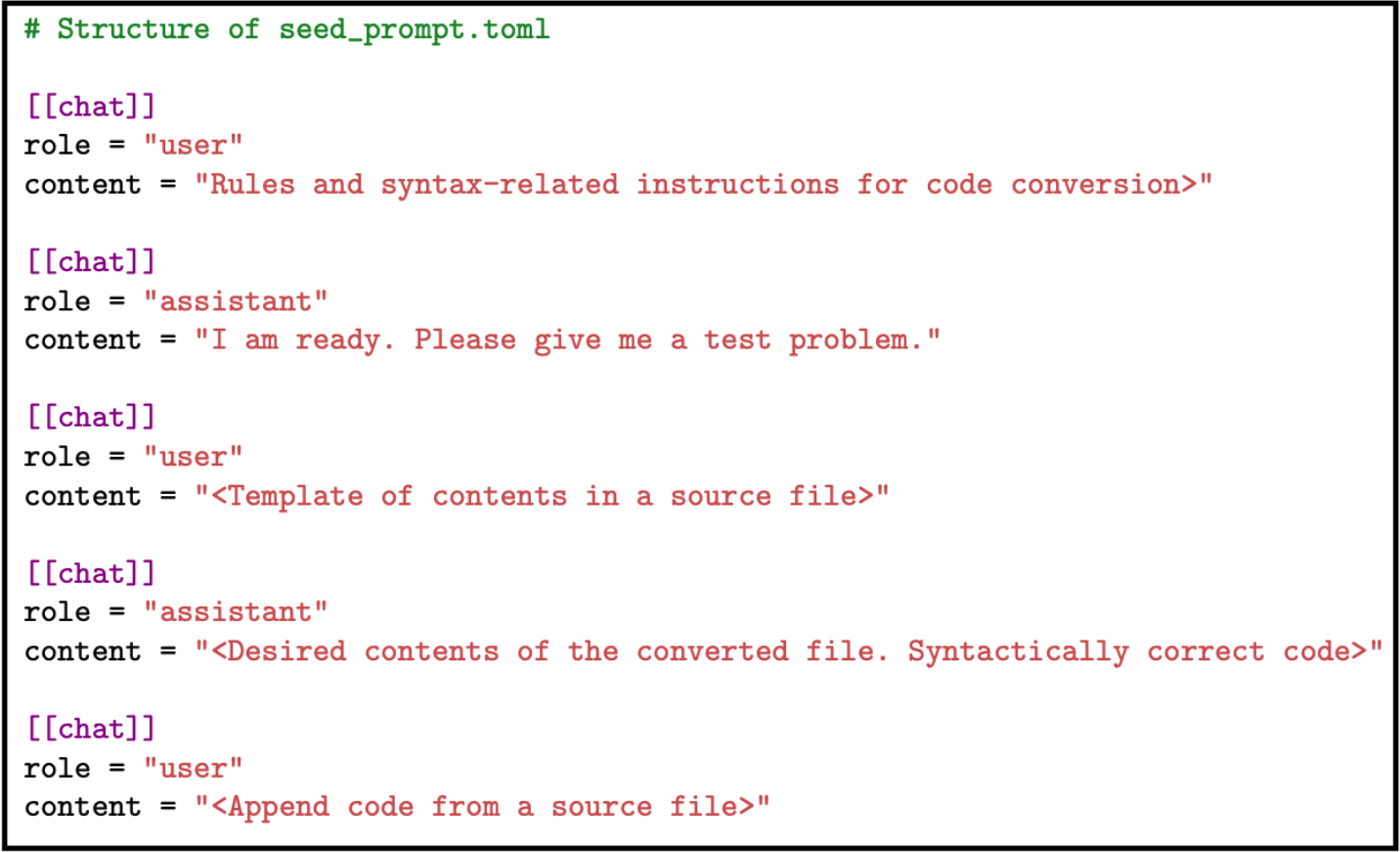}
        \caption{Structure of the chat completion template encoded in TOML format. This template outlines code conversion rules and includes an example representing a set of files that follow a similar pattern. The chat template is appended with code from the target source file and provided to the LLM, which generates the corresponding C++ and Fortran-C interface source code.}
        \Description{Structure of the chat completion template encoded in TOML format. This template outlines code conversion rules and includes an example representing a set of files that follow a similar pattern. The chat template is appended with code from the target source file and provided to the LLM, which generates the corresponding C++ and Fortran-C interface source code}
         \label{fig:chat-template}
    \end{center}
    \begin{center}
        \includegraphics[width=0.6\linewidth]{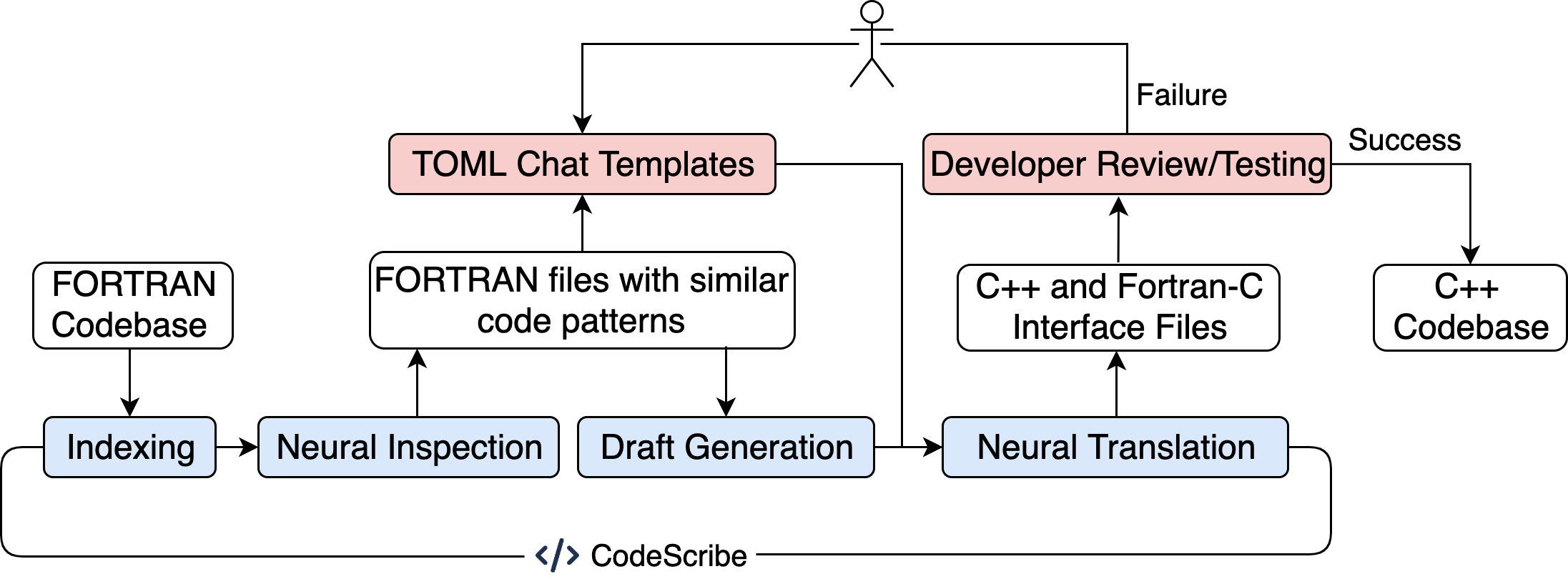}
        \caption{Schematic of the workflow for LLM-based code conversion process. Steps in blue are managed using \codescribe~while steps in red are manual and require developer intervention.}
        \Description{Schematic of the workflow required for a systematic code conversion process. Steps in blue are managed using \codescribe~while steps in red are manual and require developer intervention}
         \label{fig:llm-engine}
    \end{center}
\end{figure*}

\codescribe~operates within a systematic framework that uses LLMs to aid developers in managing the complexities of code conversion. It uses chat completion as a prompting technique to generate desired coding output from GenAI models. 
This interaction-based strategy achieves good results through structured conversations that guide the code generation process. This interaction forms the foundation of \codescribe{'s} methodology, creating templates to streamline AI-assisted code conversion.

Figure \ref{fig:chat-template} illustrates the structure of the chat template. Conversion begins by identifying patterns in the original Fortran code that are used as guide for creation of chat templates. The templates serve as references in translating different sets of files with similar source code patterns. 

Initially, the tool maps the project structure by indexing subroutines, modules, and functions across various files, which is critical for understanding the relationships between components in the codebase. Once the project structure is mapped, \codescribe~generates a draft of the C++ code for a given Fortran source file. It employs pattern recognition to replace variable declarations, include appropriate header files, and identify the use of external modules and subroutines. This information is encoded in the draft to help LLMs better understand the code context. Using the chat template as a seed prompt, \codescribe~appends the source and draft files and triggers the LLM to complete the code conversion. The generated results are extracted, reviewed, compiled, and tested for correctness. Any errors found during compilation are addressed manually by the developer or sent for regeneration by updating the seed prompt. While \codescribe~automates many aspects of the translation process, human expertise remains essential for final review and correctness, ensuring the translated code adheres to scientific computing standards. 

Figure \ref{fig:llm-engine} provides an overview of the workflow implemented by \codescribe. The schematic illustrates the systematic code conversion process, distinguishing between automated and manual steps. Blue steps indicate tasks handled by \codescribe, such as project indexing, draft generation, and LLM-driven translations, while red steps indicate manual activities, including chat template generation, code review, and verification.

\section{Methodology}
\label{sec:methodology}
Based on the overview in the previous section and the schematic in Figure \ref{fig:llm-engine}, we developed a command-line interface for \codescribe~with four commands: (1) Index, (2) Inspect, (3) Draft, and (4) Translate. Each command incorporates elements of RAG, enhancing the effectiveness of the translation process. Detailed documentation for the usage of these commands can be found in the \codescribe~repository \cite{akash_dhruv_2024_13879406}; here, we focus on providing an overview of their functionality.

\begin{figure}[h]
    \begin{center}
        \includegraphics[width=0.5\linewidth]{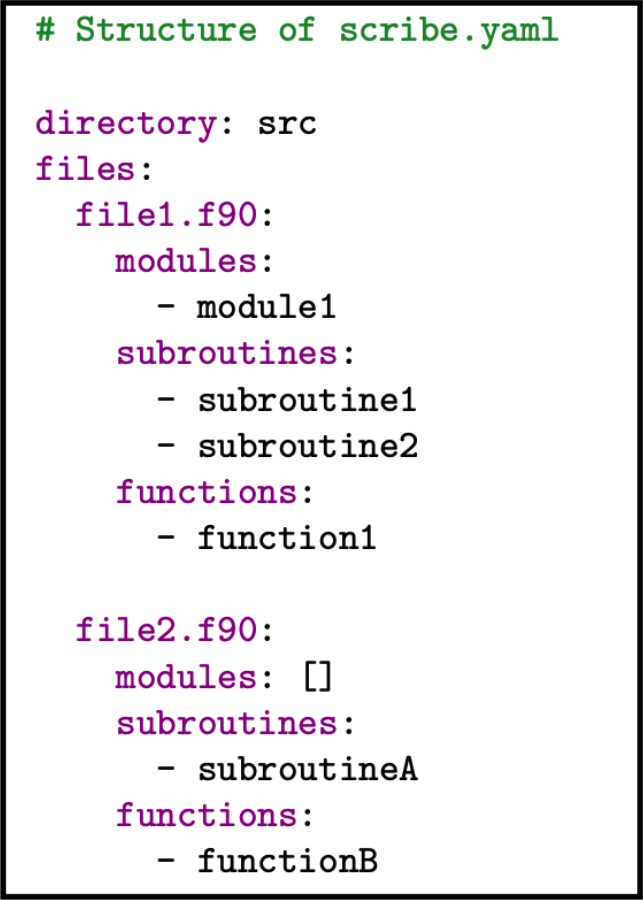}
        \caption{Sample contents of the hierarchical index file generated in each project subdirectory. This index captures relative directory location within the project structure and lists files along with associated constructs, such as modules, subroutines, and functions.}
        \Description{Sample contents of the hierarchical index file generated in each project subdirectory. This index captures relative directory location within the project structure and lists files along with associated constructs, such as modules, subroutines, and functions}
        \label{fig:scribe-yaml}
    \end{center}
\end{figure}

\subsection{Index} 
This command generates a comprehensive mapping of the source tree by analyzing the project’s file hierarchy, dependencies, and code relationships. It creates a YAML file within each subdirectory, capturing critical metadata such as the relative directory location, file names, and associated constructs like modules, subroutines, and functions. This structured index is essential for navigating large legacy codebases and ensures that subsequent translation steps are both efficient and accurate. The design of the Index command embodies RAG principles by enabling efficient retrieval of relevant metadata, which aids in informed querying during later commands. Figure \ref{fig:scribe-yaml} illustrates an example of the YAML index file. The file functions as a dictionary, where the key \codetext{directory} stores the relative path of a subdirectory to the project root defined by the key \codetext{root}. The key \codetext{files} contains a hierarchy of dictionaries with keys corresponding to filenames and constructs. For example: \codetext{[files][file1.f90][modules]} provides a list of modules in the file \codetext{file1.f90}. During the execution of other \codescribe~commands, YAML files from all directories are loaded and compiled into an inverse dictionary that maps constructs back to their corresponding filenames for efficient querying. For instance, during draft code generation, the inverse dictionary is utilized to determine that \codetext{subroutineA} belongs to \codetext{file2.f90}.

{The need for the Index command arose from observing that, without specific guidance, LLMs often attempt to fill gaps in context by generating peripheral code, particularly when subroutines or modules appear undefined in the target file. For instance, when prompted to translate a subroutine that references external functions, the LLM would often extrapolate code for these external functions or even provide a main program example to demonstrate functionality. This was especially problematic in large codebases with deeply interconnected files. To mitigate such hallucinations, we developed the Index command to capture a detailed structural map of the source code using YAML files. This layout serves as a reference point for prompts, helping the LLM to focus solely on the intended conversion without producing code for elements already defined elsewhere in the project.}

\subsection{Inspect} 
The Inspect command allows users to interactively query specific details of the source code, using an LLM to answer queries for a list of Fortran files. This feature helps developers understand the structure and behavior of individual code sections, which is particularly useful when working with large or undocumented legacy codebases. Parsing scripts extract module names, functions, and subroutines, which are then queried against the YAML index to provide context for the LLM. The command leverages RAG elements by retrieving context from the YAML index. Interaction with the LLM is achieved through one of the following mechanisms:

\begin{enumerate}
    \item Directly by sending requests to API endpoints, such as for OpenAI models \cite{openaiapi}.
    \item Locally using downloaded models via the Transformers library \cite{huggingface}, for models like Codellama and Mistral.
    \item Saving the prompt contents to a JSON file, which can then be copy/pasted into interfaces like ChatGPT for interactive discussions.
\end{enumerate}

{The Inspect command ensures that the LLM can accurately answer developer queries without over-generating responses that diverge from the target function’s scope. It not only aids in code comprehension but also facilitates knowledge generation to design the seed prompt (chat template) shown in Figure \ref{fig:chat-template} for code translation.}

\subsection{Draft} 
The Draft command automates the initial phase of code conversion by identifying patterns within the Fortran source code and applying systematic substitutions to produce a preliminary C++ draft. This step serves as preparation for AI-driven translation by generating additional context alongside the seed prompt.

{The creation of the Draft command was informed by challenges encountered in maintaining the consistency of data structures and variable types. Without a preliminary draft, LLMs frequently introduced erroneous data types or redefined existing structures, leading to incoherence in generated C++ code. Providing a draft code serves as a controlled data point, helping the LLM to align its output with the expected types and structures. Additionally, comments embedded in the draft help prevent common hallucinations, such as re-declaring functions or subroutines already defined elsewhere, by clearly indicating these constructs in advance.}

The process converts basic variable types and multidimensional arrays, and uses the construct-query mechanism for YAML files to identify modules, functions, and subroutines used from elsewhere in the source code. It also inserts comments into the draft for LLM to parse. For example, if a Fortran file \codetext{<filename>.f90} contains a scalar variable declaration \codetext{real::funcvar} but uses it as a function \codetext{funcvar(i,j)}, it may indicate that the variable is defined elsewhere in the code as a function or defined explicitly in the file as a statement function.  In such cases, the LLM is given enough context to either reference the function from an appropriate header file or generate a corresponding C++ lambda function. Similarly, when modules are imported via a \codetext{use} statement or subroutines are called with a \codetext{call} statement, the draft includes comments to guide the LLM in correctly handling these constructs. The draft code is saved with to \codetext{<filename>.scribe} file.

\subsection{Translate} 
The core functionality of \codescribe~lies in its AI-powered translation. This command transforms Fortran source code into C++ while generating the necessary Fortran-C interfaces for smooth integration between legacy and modern codebases. The translation leverages generative AI to interpret Fortran’s syntax and semantics. The seed prompt from the chat template is appended with code from the source and draft files and sent to the AI for chat completion. The source code from the Fortran file is supplied within the elements \codetext{<source>...</source>}, and the draft code is enclosed within \codetext{<draft>...</draft>}. The LLM is instructed to produce output for the C++ source and the corresponding Fortran-C interface, enclosing them within \codetext{<csource>...</csource>} and \codetext{<fsource>...</fsource>}. This mechanism is based on the implementation by the Slack team in their recent article \cite{slack}, allowing for source code extraction and saving to separate files named \codetext{<filename>.cpp} and \codetext{<filename>\_fi.f90}.

Interaction with the LLM is achieved using one of the three mechanisms employed in the Inspect command. Utilizing direct API endpoints or the Transformers library enables automated extraction and creation of C++ and Fortran interface files, while the interactive method of using JSON files with ChatGPT requires manual copy-pasting of relevant source code into its respective file.

\section{Target Codes} \label{sec:target-codes}

We explored the application of \codescribe~to the following scenarios within scientific computing:
\begin{enumerate}
    \item \textbf{Converting the \mcfm~code}: \mcfm~is a parton-level Monte Carlo program that provides predictions for a wide range of processes at hadron colliders \cite{Campbell_1999,Campbell_2011,Campbell_2019}. The code is primarily written in Fortran and requires complete conversion to C++ to enhance its interoperability with other codes and libraries in high-energy physics. {The codebase consists of approximately 450-500 source files, each containing 50-70 lines of code.}

    \item \textbf{Developing C++ interface for Noah-MP}: Noah-MP \cite{noahmp} is a Fortran library that models urban canopy parameterization for atmospheric simulations. Integrating it with the Energy Research and Forecasting (ERF) Model \cite{Almgren2023} is essential for enhancing the latter's applicability. However, ERF is written in C++. We utilized \codescribe~to create interfaces for this integration.
\end{enumerate}
LLM performance studies discussed in section \ref{sec:results} are dominated by \mcfm~code conversion because that was not only the motivator for the development of \codescribe~but has seen its most extensive use. As of this writing the entire code has been converted to C++. {The conversion process initially averaged 2-3 files per day before the development of the tool. After the tool was implemented, developer productivity increased to 10-12 files per day.} Below are a few additional statistics gleaned from this work.

\section{LLM Performance Studies} \label{sec:results}
{ We present two distinct studies: (1) The performance of \codescribe~with various models of different parameter counts, and (2) The influence of RAG on performance.}
\subsection{Model Sensitivity Study}
For this study, we focused on evaluating the performance of various GenAI models and providing a quantitative assessment of the developer time required to verify and test the code after LLM conversion. { For a more detailed look at the files and the code conversion process, we encourage readers to explore the supplementary materials available on GitHub \cite{results}. These resources contain the raw source files used in the analysis and the results discussed in this article. Relevant files can be accessed through the links marked with \faExternalLink. We will refer to the data repository as \codetext{<results>} throughout the paper.}
\begin{figure}[h]
    \begin{center}
        \includegraphics[width=0.75\linewidth]{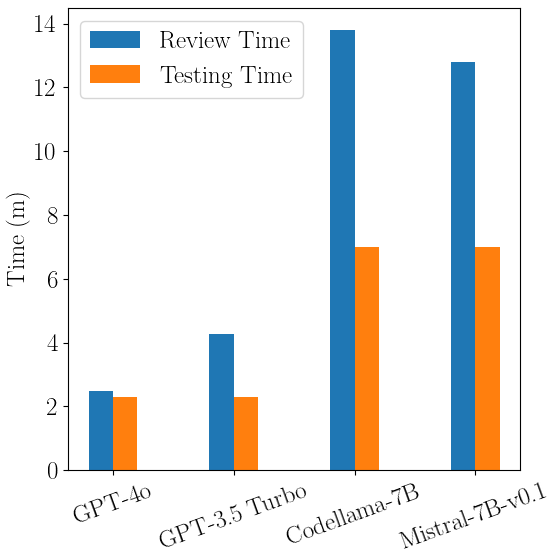}
        \caption{Developer time (minutes) for converting and testing a single Fortran source file in \mcfm~with various GenAI models.}
        \Description{Developer time (minutes) for converting and testing a single Fortran source file in \mcfm~with various GenAI models.}
         \label{fig:model-sensitivity}
    \end{center}
\end{figure}

Consider the example target Fortran file: 
\begin{itemize}
    \item[] \href{https://github.com/akashdhruv/llm-conversion-performance/blob/main/model-sensitivity/target.f}{\faExternalLink \: \codetext{<results>/model-sensitivity/target.f}}
\end{itemize}
We utilized five models for the code conversion task: CodeLlama-7B, Mistral-7B, CodeLlama-34B, GPT-4o, and GPT-3.5 Turbo, each with distinct architectures, parameter counts, and capabilities. CodeLlama-7B and CodeLlama-34B, based on Meta’s LLaMA architecture \cite{Codellama2024}, are optimized for coding tasks, differing primarily in parameter size (7 billion for CodeLlama-7B and 34 billion for CodeLlama-34B). Mistral-7B \cite{jiang2023mistral7b}, another open-source model with 7 billion parameters, employs a dense attention mechanism, achieving efficiency comparable to larger models. A key advantage of these models is their capability to run locally using organizational resources; for instance, we executed them on NVIDIA A100 GPUs using the Swing system at Argonne \cite{swing}.

GPT-4o and GPT-3.5 Turbo, developed by OpenAI, are general-purpose models accessible via API endpoints, requiring the purchase of API credits. GPT-4o is a multimodal model capable of processing both text and images, and although OpenAI has not disclosed its architecture or parameter count, it is widely speculated to have significantly more parameters than other models. GPT-3.5 Turbo is smaller and optimized for cost-effective, high-speed text generation. While GPT-4o is well-suited for complex multimodal tasks and excels in multi-step reasoning, CodeLlama and Mistral are specialized for coding tasks and do not support multi-step reasoning to the same extent as GPT-4o. {This distinction is critical when evaluating model performance in specific use cases, such as the MCFM code conversion, where the ability to engage in multi-step reasoning can significantly affect the quality of the translated code}.

Figure \ref{fig:model-sensitivity} provides an overview of the performance of different models in terms of developer time (in minutes) for converting and testing a single Fortran source file in MCFM to its corresponding C++ source and Fortran-C interface. The raw source files for each model's output, along with the reference solution, can be found in the data repository at following locations:
\begin{itemize}
    \item[] \href{https://github.com/akashdhruv/llm-conversion-performance/blob/main/model-sensitivity/codellama-7b}{\faExternalLink \: \codetext{<results>/model-sensitivity/codellama-7b}}
    \item[] \href{https://github.com/akashdhruv/llm-conversion-performance/blob/main/model-sensitivity/mistral-7b}{\faExternalLink \: \codetext{<results>/model-sensitivity/mistral-7b}}
    \item[] \href{https://github.com/akashdhruv/llm-conversion-performance/blob/main/model-sensitivity/gpt-3.5-turbo}{\faExternalLink \: \codetext{<results>/model-sensitivity/gpt-3.5-turbo}}
    \item[] \href{https://github.com/akashdhruv/llm-conversion-performance/blob/main/model-sensitivity/gpt-4o}{\faExternalLink \: \codetext{<results>/model-sensitivity/gpt-4o}}
    \item[] \href{https://github.com/akashdhruv/llm-conversion-performance/blob/main/model-sensitivity/reference}{\faExternalLink \: \codetext{<results>/model-sensitivity/reference}}
\end{itemize}
And the corresponding chat completion template is stored in:
\begin{itemize}
    \item[] \href{https://github.com/akashdhruv/llm-conversion-performance/blob/main/model-sensitivity/seed_prompt.toml}{\faExternalLink \: \codetext{<results>/model-sensitivity/seed\_prompt.toml}}
\end{itemize}
All models operated under a maximum token limit of 4096, with CodeLlama and Mistral models utilizing a maximum batch size of 8 and default values for temperature and top-p parameters.
\begin{figure*}[h]
    \begin{center}
        \includegraphics[width=0.9\linewidth]{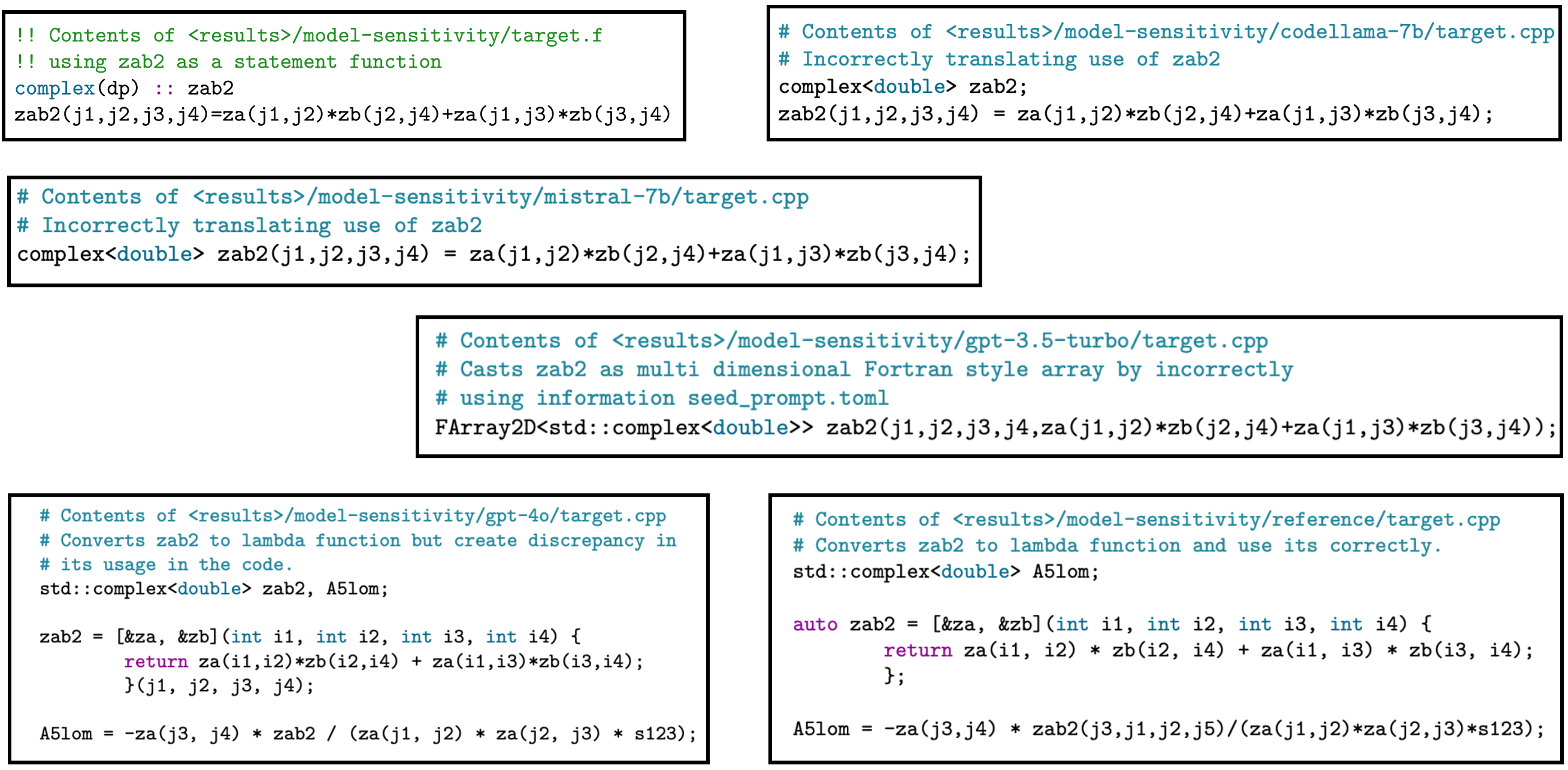}
        \caption{Excerpts from files converted using different models, highlighting the improper conversion of statement functions to lambda functions. Situations like these require manual developer intervention to ensure code correctness.}
        \Description{Excerpts from files converted using different models, highlighting the improper conversion of statement functions to lambda functions. Situations like these require manual developer intervention to ensure code correctness.}
         \label{fig:statement-function-conversion}
    \end{center}
\end{figure*}

The results reveal distinct strengths and limitations among the models. CodeLlama-7B did not generate the necessary Fortran interface code and failed to enclose the code within the specified \codetext{<csource>} and \codetext{<fsource>} elements. Additionally, it struggled to separate the main implementation from the extern C interface in the C++ source and could not convert the statement function \codetext{zab2} to a lambda function. As a result, integrating the generated files with the rest of the codebase took a total of 13.8 minutes for review and 7 minutes for testing.

In contrast, Mistral-7B exhibited better structural performance, successfully placing the code in the appropriate elements and separating the main implementation from the extern C interface. However, it hallucinated the Fortran interface by copying code directly from the \codetext{target.f} file and, like CodeLlama-7B, failed to convert the statement function \codetext{zab2}. Additionally, it performed an incomplete conversion for the variable \codetext{Fcc}. The developer time required for Mistral-7B was comparable to CodeLlama-7B, with 12.8 minutes spent in review and 7 minutes in testing.

GPT-3.5 Turbo demonstrated strong overall performance, meeting the requirements outlined in the chat template by generating the Fortran interface, the extern C interface, and the main implementation. This model notably improved review and testing times but failed to properly convert the statement function \codetext{zab2} into a lambda function. Instead, it incorrectly transformed \codetext{zab2} into a multidimensional array, misinterpreting the instructions provided in \codetext{seed\_prompt.toml}. Despite this issue, it required only 4.26 minutes in review and 2.3 minutes in testing, reflecting how better adherence to the template reduced overall time spent on manual adjustments and testing iterations.

GPT-4o slightly outperformed GPT-3.5 Turbo by following the conversion rules for the statement function; however, the conversion of \codetext{zab2} still resulted in inconsistencies in its usage throughout the C++ source file, requiring manual intervention. The review time for this model was 2.5 minutes, slightly better than GPT-3.5 Turbo, but the testing time remained comparable.

Figure \ref{fig:statement-function-conversion} provides excerpts from files converted by different models and compares them to the reference solution. These findings underscore the varying capabilities of each model in handling the MCFM code conversion, providing insights into which GenAI approaches may be most effective for software development tasks within scientific computing. GPT-4o demonstrated the best overall performance, positioning it as well-suited for the code conversion task. However, the time spent on code review was nearly equivalent to the time required for testing, suggesting inefficiencies that must be managed to reduce developer overhead. To mitigate this, we often batch the translation, review, and testing of multiple files simultaneously, optimizing the overall process and helping manage developer costs more effectively.

{ One of the primary challenges in this translation process is the handling of array indexing. Fortran uses 1-based indexing, while C++ utilizes 0-based indexing, which can be particularly tricky when dealing with complex data structures and indirect addressing schemes such as Compressed Sparse Row (CSR) indexing. Converting between 1-based and 0-based indexing is one of the most error-prone aspects of translating between these languages, especially when it impacts index values and array bounds.

To address this challenge, we developed a set of custom container classes—\codetext{FArray1D}, \codetext{FArray2D}, \codetext{FArray3D}, and so on. These classes cast C++ pointers to arrays with 1-based indexing, effectively emulating the behavior of Fortran-style arrays. The container classes are designed to allow access patterns that are very similar between Fortran and C++, and they are accessed via a special operator, ensuring a consistent array handling experience across both languages. While the initialization of arrays differs between Fortran and C++, the core array management remains consistent.

For a detailed explanation of these container classes and their usage, please refer to the raw source code repository available in the supplementary materials. Specifically, the container classes are located in:
\begin{itemize}
    \item [] \href{https://github.com/akashdhruv/llm-conversion-performance/blob/main/array-indexing/FArray.hpp}{\faExternalLink \: \codetext{<results>/array-indexing/FArray.hpp}}
\end{itemize}
An example of their application can be found in: 
\begin{itemize}
    \item[] \href{https://github.com/akashdhruv/llm-conversion-performance/blob/main/model-sensitivity/reference/target.cpp}{\faExternalLink \: \codetext{<results>/model-sensitivity/reference/target.cpp}}
\end{itemize}
(lines 56–57). These resources demonstrate the approach in full and offer insight into the indexing strategies employed to maintain consistency between Fortran and C++.}
\begin{figure}[h]
    \begin{center}
        \includegraphics[width=\linewidth]{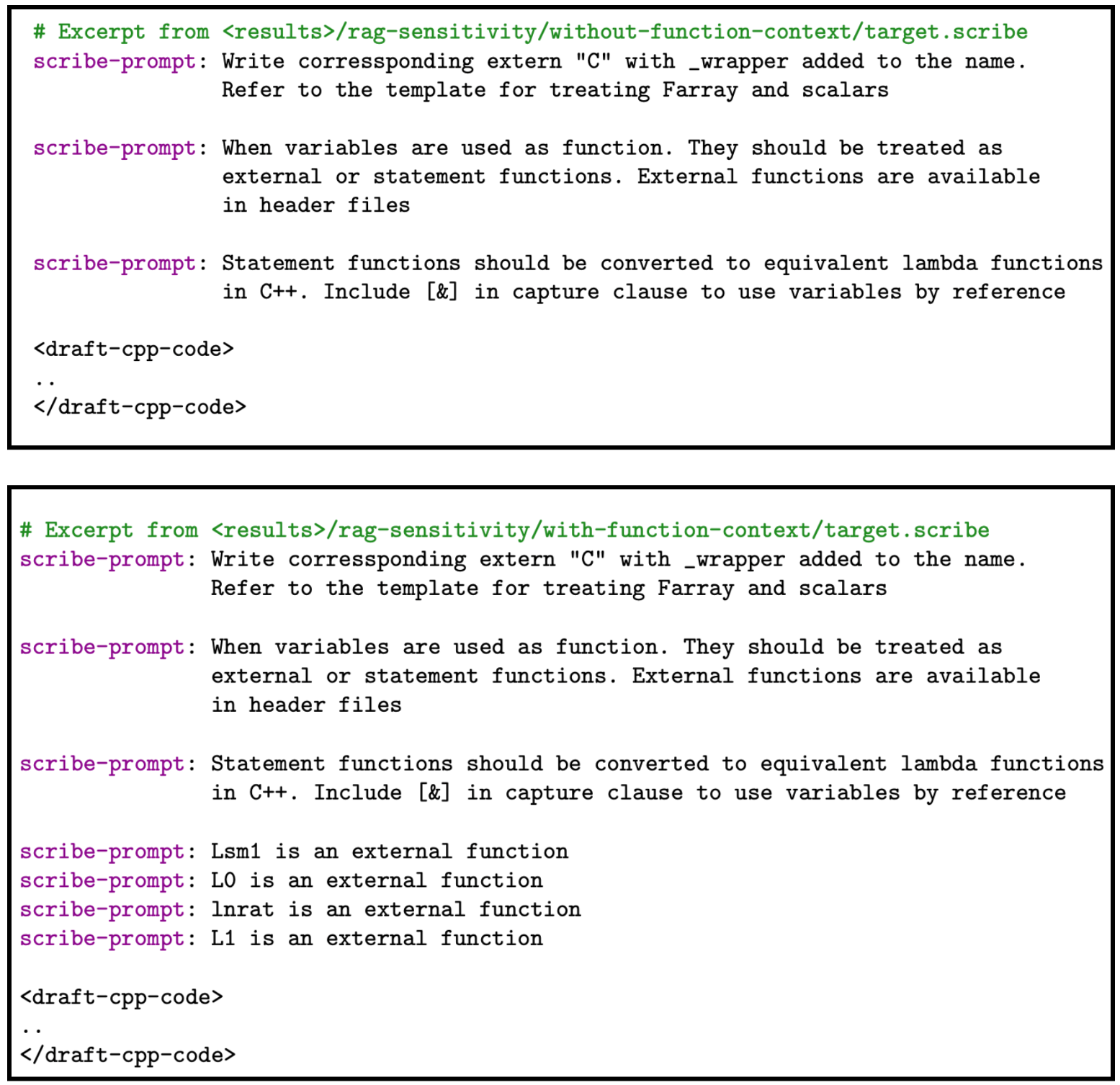}
        \caption{Excerpts from draft files used to provide LLMs with initial C++ code, project structure, and context about code conversion nuances, including the use of external functions.}
        \Description{Excerpts from draft files used to provide LLMs with initial C++ code, project structure, and context about code conversion nuances, including the use of external functions.}
         \label{fig:scribe-files-rag}
    \end{center}
%
%
    \begin{center}
        \includegraphics[width=\linewidth]{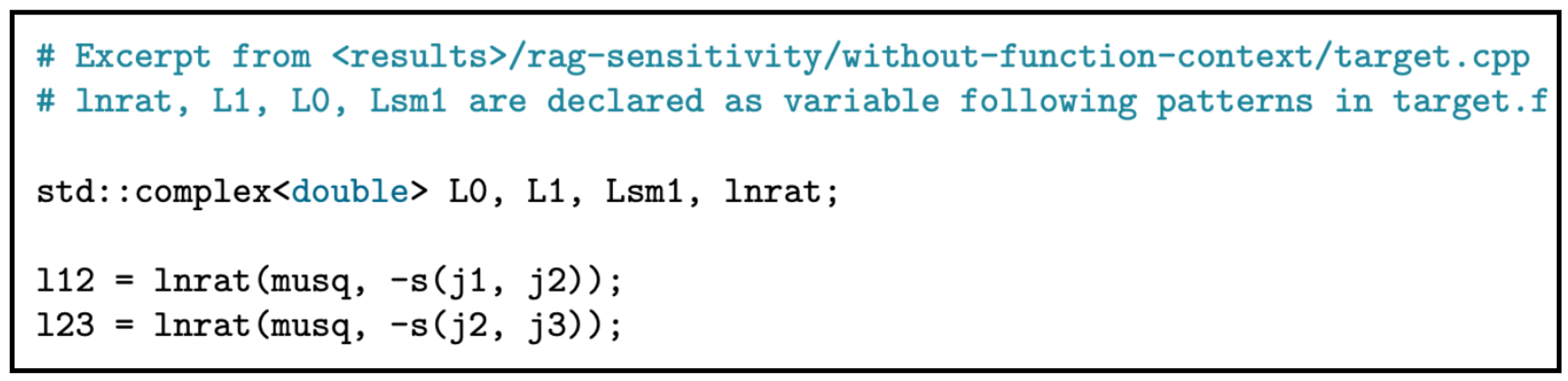}
        \caption{Excerpt from C++ code generated without context regarding the use of external functions.}
        \Description{Excerpt from C++ code generated without context regarding the use of external functions.}
         \label{fig:cpp-file-rag}
    \end{center}
\end{figure}

\subsection{RAG Sensitivity Study}

In this study, we focus on how retrieval-augmented generation (RAG) enhances code translation performance. RAG has previously been shown to improve the accuracy of natural language processing (NLP) tasks, as demonstrated by Lewis et al. \cite{lewis2020retrieval}. To evaluate its effectiveness in code translation, we applied RAG elements within \codescribe, specifically during the Draft command. In this step, a \codetext{target.scribe} file is created, and RAG is employed by parsing YAML index files and annotating the draft with prompts that provide context to the LLM, particularly about the presence of external functions.

Figure \ref{fig:scribe-files-rag} and \ref{fig:cpp-file-rag} present excerpts from two different scribe files, with and without RAG elements, alongside AI-generated C++ code. These files are available in the \codetext{results} repository:
\begin{itemize}
    \item[] \href{https://github.com/akashdhruv/llm-conversion-performance/blob/main/rag-sensitivity/with-function-context}{\faExternalLink \: \codetext{<results>/rag-sensitivity/with-function-context}}
    \item[] \href{https://github.com/akashdhruv/llm-conversion-performance/blob/main/rag-sensitivity/without-function-context}{\faExternalLink \: \codetext{<results>/rag-sensitivity/without-function-context}}
\end{itemize}

The results demonstrate that when provided with the context of external functions--\codetext{lnrat}, \codetext{L0}, \codetext{L1}, \codetext{Lsm1}--the AI is able to detect their use and avoid declaring them explicitly. Without this context, however, the functions are declared as they appear in the \codetext{target.f} file, found at:
\begin{itemize}
    \item[] \href{https://github.com/akashdhruv/llm-conversion-performance/blob/main/rag-sensitivity/target.f}{\faExternalLink \: \codetext{<results>/rag-sensitivity/target.f}}
\end{itemize}

In Fortran, external functions can be declared as variables, and the compiler identifies them during linkage. However, in C++, this leads to linkage errors. While the RAG-enhanced draft file annotations improve accuracy, the results are inconsistent, as the LLM occasionally hallucinates despite being explicitly informed of the use of external functions. Our ongoing efforts focus on mitigating these hallucinations by implementing a more robust RAG framework.

\section{Ongoing Work} \label{sec:ongoing-work}

In our ongoing work, we are applying the code conversion techniques developed for \mcfm~to design a Fortran-C interoperability layer for Noah-MP and ERF, with the goal of incorporating Urban Canopy parameterization into ERF simulations. { Use of \codescribe~for Noah-MP demonstrates a systematic approach to type conversion from Fortran to C++ through modern interoperability features. The code leverages Fortran derived types defined within modular structures using constructs introduced in Fortran 2003—with selective use of Fortran 2008 features—and interfaces them with C++ via the iso\_c\_binding intrinsic. This guarantees that memory layouts remain consistent between the languages. In C++, mirror structures are created to directly reference Fortran-managed memory, with the implementation relying on features available from C++11 onwards. The conversion process, carried out through dedicated initialization routines, maps Fortran arrays and pointers to their C++ counterparts while preserving Fortran’s indexing conventions, thereby supporting robust and efficient cross-language data sharing. Below in an example of Noah-MP type conversion:
\begin{itemize}
    \item[] \href{https://github.com/akashdhruv/llm-conversion-performance/blob/main/type-matching}{\faExternalLink \: \codetext{<results>/type-matching}}
\end{itemize}
The example demonstrates how \codetext{NoahmpIO\_type} is mirrored between Fortran and C++ through following implementation:
\begin{itemize}
    \item[] \href{https://github.com/akashdhruv/llm-conversion-performance/blob/main/type-matching/NoahmpIOVarType.F90}{\faExternalLink \: \codetext{<results>/type-matching/NoahmpIOVarType.F90}}
    \item[] \href{https://github.com/akashdhruv/llm-conversion-performance/blob/main/type-matching/NoahmpIO.H}{\faExternalLink \: \codetext{<results>/type-matching/NoahmpIO.H}}
     \item[] \href{https://github.com/akashdhruv/llm-conversion-performance/blob/main/type-matching/NoahmpIO.cpp}{\faExternalLink \: \codetext{<results>/type-matching/NoahmpIO.cpp}}
\end{itemize}
}



To enhance developer productivity during code conversion, we plan to expand the Inspect command to automatically generate \codetext{seed\_prompt.toml} chat templates and to integrate \codescribe~with the LASSI framework \cite{Dearing2024} to improve the correctness of the generated code.

\section{Conclusions}
In this study we presented \codescribe, a GenAI-driven tool designed to facilitate the translation of legacy Fortran code to modern C++ while creating necessary interfaces for seamless integration. Our evaluations demonstrate that \codescribe~reduces the developer time required for code conversion and testing, showcasing its potential to enhance productivity in scientific computing projects.

We explored the effectiveness of various AI models for code conversion of the MCFM codebase, revealing distinct strengths and weaknesses across different architectures. While GPT-4o emerged as the most effective model in this context, the performance of GenAI in code conversion also revealed opportunities for optimization, particularly concerning the manual review and testing processes that often accompany such translations.

In addition to the MCFM code conversion, our ongoing efforts focus on applying \codescribe~to meet developer requirements across various scientific computing projects, such as Noah-MP and Flash-X, by enhancing existing commands and introducing new functionality. { Flash-X \cite{DUBEY2022101168} is an open-source multiphysics simulation software designed for modeling incompressible multiphase fluid dynamics with heat transfer. It is widely used in the simulation of pool and flow boiling phenomena, particularly in phase-change cooling systems \cite{Dhruv_2024}. Flash-X incorporates AMReX \cite{Amrex_JOSS} for Adaptive Mesh Refinement (AMR) grid support. However, it currently only supports the AMReX Fortran interface, which limits its compatibility with GPUs. Using \codescribe~to build GPU compatibility between Flash-X and AMReX would significantly enhance the performance and scalability of these simulations.}

As we continue to refine \codescribe~and explore its applications, our aim is to further streamline the translation process and reduce overhead associated with code review and testing by exploring its integration with the LASSI framework developed at Argonne. We envision \codescribe~as a valuable tool that empowers developers in scientific computing to leverage GenAI effectively. 

\section{Acknowledgements}
The submitted manuscript was created in part by UChicago Argonne, LLC,
operator of Argonne National Laboratory (“Argonne”). Argonne, a
U.S. Department of Energy Office of Science laboratory, is operated
under Contract No. DE-AC02-06CH11357. The U.S. Government retains for
itself, and others acting on its behalf, a paid-up nonexclusive,
irrevocable worldwide license in said article to reproduce, prepare
derivative works, distribute copies to the public, and perform
publicly and display publicly, by or on behalf of the Government.  The
Department of Energy will provide public access to these results of
federally sponsored research in accordance with the DOE Public Access
Plan. http://energy.gov/downloads/doe-public-access-plan.

The City of Chicago is located on land that is and has
long been a center for Native peoples. The area is the traditional homelands of
the Anishinaabe, or the Council of the Three Fires: the Ojibwe, Odawa, and Potawatomi Nations.
Many other Nations consider this area their traditional homeland, including the Myaamia, Ho-Chunk, Menominee, Sac and Fox, Peoria, Kaskaskia, Wea, Kickapoo, and Mascouten.

\bibliographystyle{ACM-Reference-Format}
\bibliography{biblio}

\end{document}